\documentclass[aps,pre,showpacs,amsmath,amssymb,amsfonts,superscriptaddress,lengthcheck,twocolumn,longbibliography]{revtex4-1}

\usepackage{color}
\usepackage{graphicx}
\usepackage{amsthm}
\usepackage{subfigure}
\usepackage{tabu}
\usepackage{verbatim}
\usepackage{dcolumn}
\usepackage{bm}
\usepackage{epsf}
\usepackage{color}
\usepackage[colorlinks=true,citecolor=blue,linkcolor=blue,urlcolor=blue]{hyperref}
\usepackage{hhline}
\usepackage{mathtools}
\usepackage{bbold}

\newcommand{\be}{\begin{equation}}
\newcommand{\ee}{\end{equation}}
\newcommand{\ba}{\begin{align}}
\newcommand{\ea}{\end{align}}
\newcommand{\bi}{\begin{itemize}}
\newcommand{\ei}{\end{itemize}}

\newcommand{\bla}{bla\\bla\\bla\\bla\\bla}

\begin{document}

\title{Kibble-Zurek mechanism in driven underdamped Brownian motion}

\author{Pierre Naz\'e}
\email{pierre.naze@icen.ufpa.br}

\affiliation{\it Universidade Federal do Par\'a, Faculdade de F\'isica, ICEN,
Av. Augusto Correa, 1, Guam\'a, 66075-110, Bel\'em, Par\'a, Brazil}

\date{\today}

\begin{abstract}

Kibble-Zurek mechanism is widely known to appear in the transverse-field quantum Ising chain in the thermodynamic limit at zero temperature, having notorious characteristics, like the divergence of its relaxation time. In this work, I present the same effect in a simple system, the driven underdamped Brownian motion. Using linear response theory, I show the appropriate limits where the Kibble-Zurek mechanism happens. The divergence of the relaxation time, the high-temperature condition in the initial thermal equilibrium state, a new Kibble-Zurek scaling at sudden processes, and the pausing effect in the optimal protocol are presented as consequences.

\end{abstract}

\maketitle




\section{Underdamped Brownian motion}

Consider an underdamped Brownian motion driven in the stiffening parameter, according to the Langevin equation~\cite{gomez2008optimal}
\begin{equation}
    m \ddot{x}(t)+\gamma \dot{x}(t)+\omega_0^2(t)x(t)=\eta(t),
\end{equation}
where $m$ is the mass of the particle, $\gamma$ is the friction parameter, $\omega_0(t)$ the time-dependent natural frequency and $x(t)$ the position of the particle. The random force $\eta(t)$ is a white noise, such that
\begin{equation}
    \langle \eta(t)\rangle=0,\quad \langle \eta(t)\eta(t')\rangle=\frac{2m\gamma}{\beta}\delta(t-t'),
\end{equation}
where $\beta$ is proportional to the inverse initial temperature $T$. Using linear response theory~\cite{kubo2012statistical}, the relaxation function is given by
\begin{multline}
\frac{\Psi_0(t)}{\Psi_0(0)}=e^{-\gamma|t|}\left[2+\left(\frac{\omega^2}{\omega_0^2}-2\right)\cos{\omega t}+\frac{\gamma\omega}{\omega_0^2}\sin{\omega |t|}\right],
\end{multline}
\begin{equation}
    \Psi_0(0)=\frac{1}{2m^2\beta \omega_0^2\omega^2},
\end{equation}
where $\omega=\sqrt{4\omega_0^2-\gamma^2}$ is a positive number. The relaxation time is 
\begin{equation}
    \tau_R(\gamma,\omega_0)=\frac{\gamma^2+\omega_0^2}{2\gamma\omega_0^2}.
\end{equation}
We are going to see now that the conditions to achieve Kibble-Zurek mechanism is the limits $\gamma\rightarrow 0^+$ and $\omega_0\rightarrow 0^+$, at the same rates, with $\gamma/\omega_0=1<2$. Observe that such conditions imply $\omega\ge 0$.

\subsection{Divergence of the relaxation time}

Considering initial situations where the limits $\gamma\rightarrow 0^+$ and $\omega_0\rightarrow 0^+$, at the same rates, with $\gamma/\omega_0=1$, one has
\begin{equation}
    \lim_{\underset{\gamma/\omega_0=1}{\gamma,\omega_0\rightarrow 0^+}} \tau_R(\gamma,\omega_0)=+\infty,
\end{equation}
\begin{equation}
    \lim_{\underset{\gamma/\omega_0=1}{\gamma,\omega_0\rightarrow 0^+}} \frac{\Psi_0(t)}{\Psi_0(0)}=3,
\end{equation}
\begin{equation}
    \lim_{\underset{\gamma/\omega_0=1}{\gamma,\omega_0\rightarrow 0^+}} \Psi_0(0)=+\infty.
\end{equation}
Such an example illustrates the same effects of the Kibble-Zurek mechanism: the relaxation time and the norm of the relaxation function diverge, and the relaxation function is a constant, exhibiting a ``frozen'' state of the system, where the system takes a very long time to equilibrate along the driving. Why does this happen?

\subsection{Initial high temperature}

According to linear response theory, the system must start in an initial state of thermal equilibrium. According to the fluctuation-dissipation theorem~\cite{kubo2012statistical}, the requirement necessary to achieve this equilibration is
\begin{equation}
    T\propto \frac{1}{\gamma},
\end{equation}
indicating that the initial temperature of equilibrium must be very high. In this manner, the system should start with high thermal fluctuations, indicating a similar scenario with Kibble-Zurek mechanism, where the adiabatic theorem breaks down and the system enters a non-equilibrium zone. This explains why the relaxation function is a constant: due to the high thermal fluctuations, the system does not equilibrate along this very weak driving.

\subsection{Kibble-Zurek scaling}

\begin{figure}[t]
    \centering
    \includegraphics[width=0.9\linewidth]{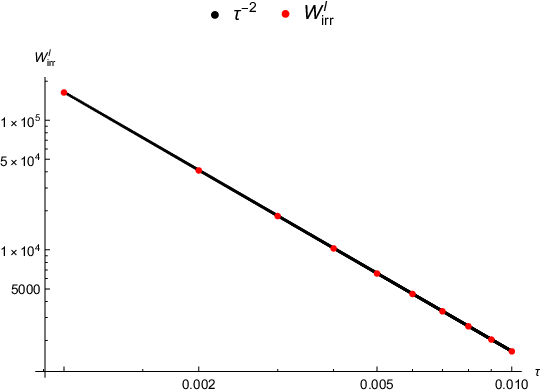}
    \caption{Kibble-Zurek scaling for sudden processes. It is given by $\eta_{KZ}=-2$, considering $\gamma=0.01$, $\omega_0=0.01$, $\delta \omega_0=0.001$, meaning that the relaxation time of the system is $\tau_R=100$.}
    \label{fig:1}
\end{figure}

Considering a linear driving, given by
\begin{equation}
    \omega(t)=\omega_0-\delta\omega_0\frac{t}{\tau},
\end{equation}
the solution of the equation where the relaxation time is equal to the rate
\begin{equation}
    \tau_R(\gamma,\omega(\hat{t}))=\hat{t}
\end{equation}
is given by $\hat{t}=f(\tau)$. Also, the irreversible work related to the impulse part is
\begin{equation}
    W_{\rm irr}^I = \frac{\delta\omega_0^2}{\tau^2}\int_{\tau-f(\tau)}^\tau\int_{\tau-f(\tau)}^t \frac{\Psi_0(t-t')}{\Psi_0(0)}dtdt'.
\end{equation}
For sudden processes, such quantity has the following scaling in the switching time
\begin{equation}
    \eta_{\rm KZ}=-2.
\end{equation}
In Fig.~\ref{fig:1}, we show a graphic considering $\gamma=0.01$, $\omega_0=0.01$, $\delta \omega_0=0.001$, meaning that the relaxation time of the system is $\tau_R=100$. Therefore, for the range of switching time used, the system performs indeed sudden processes. The stark difference with Kibble-Zurek mechanism~\cite{deffner2017kibble} comes probably from the nature of the relaxation time of the new system.

\begin{figure}[b]
    \centering
    \includegraphics[width=0.9\linewidth]{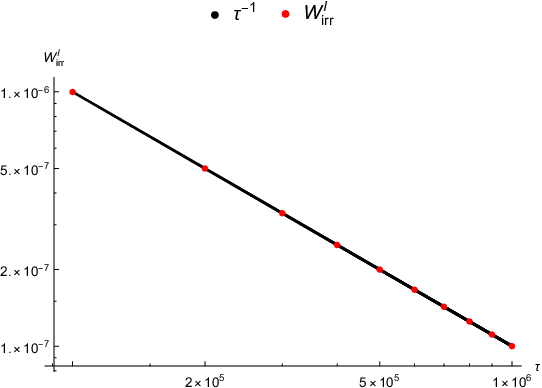}
    \caption{Kibble-Zurek scaling for slowly-varying processes. It is given by $\eta_{KZ}=-1$, considering $\gamma=0.01$, $\omega_0=0.01$, $\delta \omega_0=0.001$, meaning that the relaxation time of the system is $\tau_R=100$.}
    \label{fig:2}
\end{figure}

For slowly-varying processes, such quantity has the following scaling in the switching time
\begin{equation}
    \eta_{\rm KZ}=-1.
\end{equation}
In Fig.~\ref{fig:2}, we show a graphic considering $\gamma=0.01$, $\omega_0=0.01$, $\delta \omega_0=0.001$, meaning that the relaxation time of the system is $\tau_R=100$. Therefore, for the range of switching time used, the system performs indeed slowly-varying processes. The same exponent occurs with the traditional transverse-field quantum Ising chain~\cite{deffner2017kibble}.

Observe that it is not guaranteed that the values of the irreversible work in the impulse part correspond to the exact value. Indeed the solution $f(\tau)$ is always greater than the range of $\tau$ used in these graphics, which indicates that the range of validity of linear response is not respected. However, previous studies illustrate that the Kibble-Zurek exponent is calculated correctly~\cite{naze2022kibble}. 

\subsection{Optimal protocol: pausing effect}

When Kibble-Zurek mechanism happens, observe that $\tau/\tau_R\rightarrow 0^+$. The optimal protocol for the average work and its fluctuations, in linear response theory, will be~\cite{naze2024analytical}
\begin{equation}
    g^*(t)\approx \frac{1}{2},
\end{equation}
where the system ``pauses'' in the middle of the driving, performing two jumps, one at the beginning of the process, and another at the end. Basically, this means that the system does not interact with the environment, which is reasonable for an optimized process. Observe that the optimal protocol ``pauses'' in the critical point in the traditional Kibble-Zurek occurs because the driving is symmetric around such point~\cite{chen2020and}. It is just a coincidence.

\section{Final remarks} 
\label{sec:finalremarks}

In this work, respecting all the characteristics of the underdamped Brownian motion, I present the limit where the famous Kibble-Zurek mechanism happens. This phenomenon occurs because the system is driven with very weak stiffness from an initial high-temperature bath, and achieves a ``frozen'' state as the system takes a long time to equilibrate along the driving due to the high thermal fluctuations. I present the main consequence of this fact: divergence of its relaxation time, the initial high-temperature heat bath, a new Kibble-Zurek scaling for sudden processes, and the pausing effect for optimal protocols. Last but not least, if the initial parameters are sufficiently controlled, the underdamped Brownian motion could be a more accessible example to study the Kibble-Zurek mechanism.

\bibliography{KZUB}
\bibliographystyle{apsrev4-2}

\end{document}